\documentclass[conference,9pt]{IEEEtran}
\IEEEoverridecommandlockouts
\usepackage{amsmath,amssymb,amsfonts}
\usepackage{algorithmic}
\usepackage{graphicx}
\usepackage{textcomp}
\usepackage{xcolor}
\usepackage{booktabs}
\usepackage{multirow}
\usepackage{makecell}
\usepackage[numbers,sort&compress]{natbib}
\def\BibTeX{{\rm B\kern-.05em{\sc i\kern-.025em b}\kern-.08em
    T\kern-.1667em\lower.7ex\hbox{E}\kern-.125emX}}

\begin{document}

\title{Basket-Enhanced Heterogenous Hypergraph for Price-Sensitive Next Basket Recommendation
\thanks{\textsuperscript{\dag} These authors contributed equally to this work.

\textsuperscript{*} Corresponding Author

}
}

\author{\IEEEauthorblockN{Yuening Zhou\textsuperscript{\dag}}
\IEEEauthorblockA{\textit{Department of Marketing} \\
\textit{The Chinese University of Hong Kong}\\
Hong Kong, China \\
yueningzhou@cuhk.edu.hk}
\and
\IEEEauthorblockN{Yulin Wang\textsuperscript{\dag}}
\IEEEauthorblockA{\textit{School of Computer Science and Technology} \\
\textit{Dalian University of Technology}\\
Dalian, China \\
15683911223@163.com}
\and
\IEEEauthorblockN{Qian Cui}
\IEEEauthorblockA{\textit{The Data Intelligence Team} \\
\textit{Maoyan}\\
Beijing, China \\
brown3qqq@foxmail.com}
\and
\IEEEauthorblockN{Xinyu Guan}
\IEEEauthorblockA{\textit{The Security Platform Department} \\
\textit{Tencent}\\
Shenzhen, China \\
xinyu\_guan1998@outlook.com}
\and
\IEEEauthorblockN{Francisco Cisternas\textsuperscript{*}}
\IEEEauthorblockA{\textit{Department of Marketing} \\
\textit{The Chinese University of Hong Kong}\\
Hong Kong, China \\
fcisternas@cuhk.edu.hk}
}

\maketitle
\renewcommand{\footnoterule}{
    \hrule width \linewidth height 0.4pt
    \vspace{5pt} 
}

\begin{abstract}
Next Basket Recommendation (NBR) is a new type of recommender system that predicts combinations of items users are likely to purchase together. Existing NBR models often overlook a crucial factor, which is price, and do not fully capture item-basket-user interactions. To address these limitations, we propose a novel method called Basket-augmented Dynamic Heterogeneous Hypergraph (BDHH). BDHH utilizes a heterogeneous multi-relational graph to capture the intricate relationships among item features, with price as a critical factor. Moreover, our approach includes a basket-guided dynamic augmentation network that could dynamically enhances item-basket-user interactions. Experiments on real-world datasets demonstrate that BDHH significantly improves recommendation accuracy, providing a more comprehensive understanding of user behavior.
\end{abstract}


\section{Introduction}\label{Intro}
Recommender systems process vast data to provide personalized recommendations, enhancing user experience by offering tailored suggestions. While traditional systems mainly recommend individual items \cite{desrosiers2010comprehensive,wu2019session}, real-world shopping often involves purchasing multiple items in one transaction. This makes recommending item combinations, known as Next Basket Recommendation (NBR) \cite{rendle2010factorizing}.

A key difficulty in recommender systems is capturing the relationship between historical transactions and future recommendations. Existing NBR models can be classified into two main categories: item-level connections and basket-level connections. Item-level models focus on the relationships among items within a basket, incorporating purchase patterns such as co-purchases \cite{le2019correlation,chou2023incorporating} and repeat purchases \cite{ariannezhad2022recanet,romanov2023time,hu2020modeling}. Basket-level models \cite{bai2018attribute,yu2016dynamic,sun2023generative,faggioli2020recency}, often based on recurrent neural networks (RNNs) and attention mechanisms, capture users' sequential purchasing behavior across different transactions. Despite significant advancements, most of these models have not fully developed item-basket-user interactions and overlook a critical factor (i.e., price) in consumer decision, focusing solely on product choice.

\begin{figure}[t!]
    \centering
    \includegraphics[width=1\linewidth]{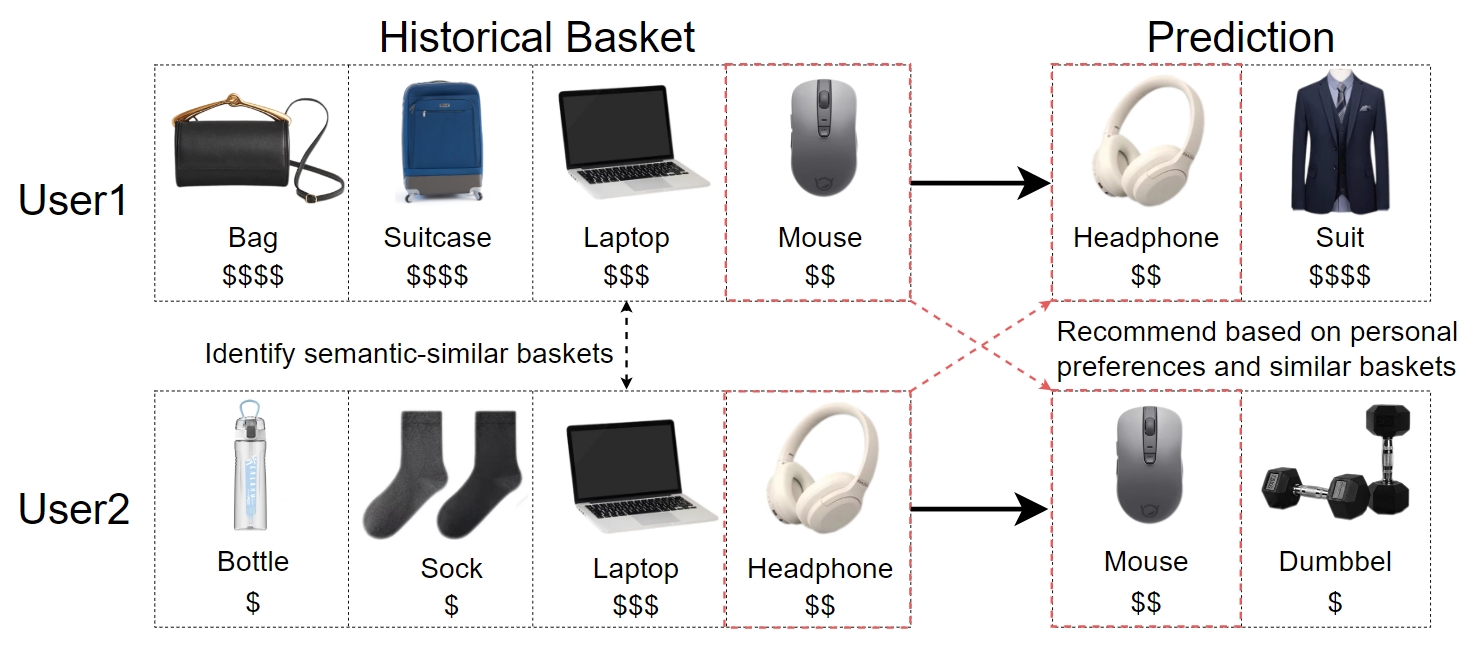}
    \caption{An example of BDHH illustration. }
    \label{fig:BDHH example}
\end{figure}

In real-world recommendation scenarios, a user's purchasing behavior is driven by their multi-aspect interests, with both product preferences and price sensitivity playing crucial roles. Price influences not only the likelihood of a purchase \cite{schafer1999recommender} but also reflect individual consumption budget. Incorporating price into recommender systems is challenging due to its complexity, which varies across consumers and product categories \cite{zheng2020price}. While some early studies have acknowledged its importance \cite{schafer1999recommender,kamishima2011personalized,chen2014does,umberto2015developing,zhang2022price}, none have specifically addressed NBR, making the first gap. Moreover, NBR considers items in a basket as interdependent, where item combinations better capture a user's overall preferences and sequences of baskets reflect evolving tastes over time. So it is desirable to identify potential products from semantically similar users by building relationships among items, baskets and users, while considering multi-aspects interest factors. Specifically, multiple items within a same basket reflect the user's broader preferences, including price sensitivity and product choices. Then the presence of the same item in different users' baskets indicates shared preferences, suggesting that items preferred by one user could be recommended to others with similar tastes, as illustrated in Fig. \ref{fig:BDHH example}. This process highlights the need for models to consider these relationships comprehensively while most existing works focus solely on item-item or basket-basket interactions, which is the second gap.

In this paper, we propose a novel method named \textbf{B}asket-augmented \textbf{D}ynamic \textbf{H}eterogeneous \textbf{H}ypergraph (BDHH) for NBR. To address the first research gap mentioned above in current research, we employ a heterogeneous multi-relational graph with different types of nodes, including item ID, price and category, and then explore the interactions between these nodes in a coarse-to-fine manner. After that, we design a unified global hybrid encoder to extract cross-associations between multiple features from the global sequence. To address the second gap, we develop a basket-guided dynamic augmentation network to establish an item-basket-item learning mechanism enhancing the relationships among items, baskets and users. This method incorporate more types of items features and achieve personalized item representation enhancement without losing global feature information. Finally, based on the semantics-enhanced node embeddings, we model user behavior through Item-ID and Item-Price hyperedges.

In summary, our contributions are as follows:
\begin{itemize}
\item We build a model considering various user interests (i.e., price sensitivity and product preference) and emphasize identifying similar semantic neighbours to connect items, baskets and users. To the best of our knowledge, we are the first to incorporate price into the NBR problem.
\item We propose a novel method called Basket-augmented Dynamic Heterogeneous Hypergraph (BDHH), which enhances user behavior representation for better performance.
\item Extensive experiments on two real-word datasets demonstrate the effectiveness of our proposed approach.
\end{itemize}

\section{Methodology}\label{Method}
\begin{figure*}[t!]
    \centering
    \includegraphics[width=1\linewidth]{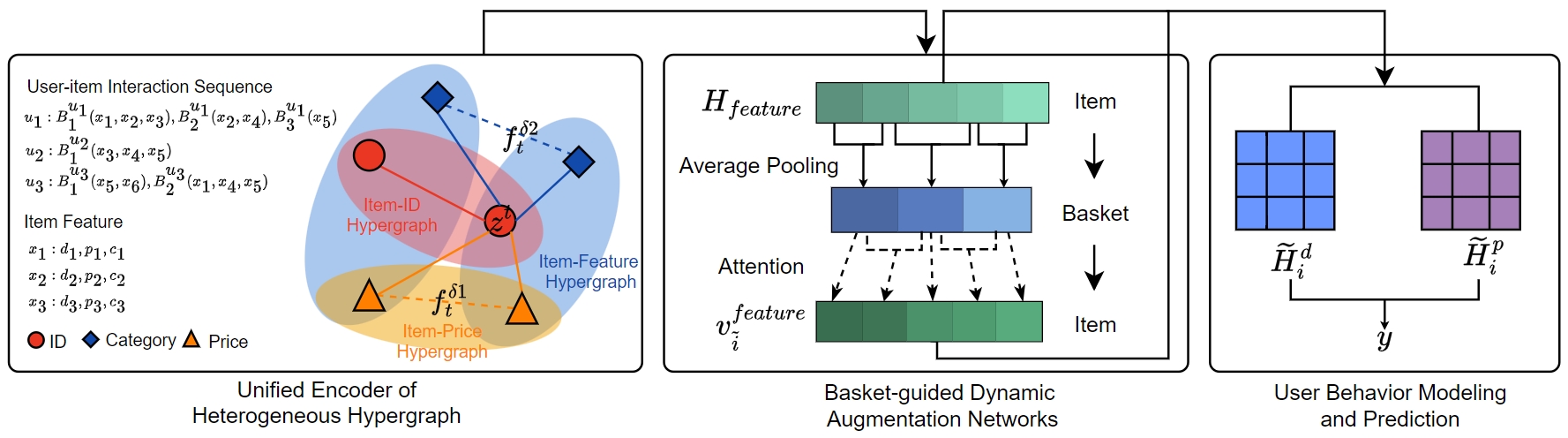}
    \caption{The framework of the proposed BDHH model.}
    \label{fig:BDHH}
\end{figure*}
In this section, we first outline the problem formulation for the NBR and then introduce our proposed model BDHH. As illustrated in Fig. \ref{fig:BDHH}, we begin by constructing a heterogeneous hypergraph network to capture item features \cite{zhang2022price,sun2021heterogeneous,linmei2019heterogeneous}. To effectively improve node representations, we design a basket-augmented mechanism, that could dynamically capture the relationships between baskets and items. Finally, we make predictions based on user's behaviors. We then describe the proposed BDHH model in detail.

\subsection{Problem Formulation}
In the NBR task, each user's purchase history is represented as a sequence of baskets $B=\{B_1^u,B_2^u,...,B_n^u\}$, where $B_i^u$ denotes the $i$-th basket of user $u\in U$. Each basket typically contains a set of items $x_i \in X$, with each item $x_i$ characterized by attributes such as item ID, price and category. The goal of the NBR task is to predict a ranked list of items for each user at a given time, specifically for the next basket $B_{n+1}^u$, based on their historical transaction data.

\subsection{Heterogeneous Hypergraph Construction Module}
In this paper, we define a heterogeneous hypergraph as $\mathcal{G} = \{ \mathcal{V}, \mathcal{E}, \mathcal{T}_{v}, \mathcal{T}_{e} \}$. Here $\mathcal{V}$ represents the set of all vertex features, including item IDs $D=\{d_1,\ldots,d_{m_d}\}$, prices $P=\{p_1,\dots,p_{m_p}\}$ and categories $C=\{c_1,\dots,c_{m_c}\}$, so $\mathcal{V}=D\cup P\cup C$. Note that we use price discretization to obtain relative price instead of absolute price according to \cite{zheng2020price}. The set of hyperedges $\mathcal{E}$, includes connections where each hyperedge $\epsilon\in \mathcal{E}$ can link multiple nodes. The graph consists of three types of hyperedges: an item-feature hypergraph connecting all features of an item, an item-ID hypergraph linking all item IDs within a basket, and an item-price hypergraph connecting all item price within a basket. $\mathcal{T}_{v}$ is the set of node features and $\mathcal{T}_{e}$ is the set of hyperedge types. Since $\lvert \mathcal{T}_{v} \rvert+\lvert \mathcal{T}_{e} \rvert>2$, the hypergraph is considered heterogeneous.

\subsection{Unified Global Hybrid Encoder}
For a specific node, different neighbouring nodes may have different impacts on it. To integrate neighbouring homogeneous nodes and leverage heterogeneous information between different types of nodes, we introduce a unified global hybrid node embedding update method. Given a specific node $z^t \in \mathbb{R}^d$ with feature $t$, we could represent the semantics-enhanced embeddings of this node as:
\begin{equation}\label{eq:h^t}
    h^t=\gamma_t\odot f_t^{\delta1}+(1-\gamma_t)\odot f_t^{\delta2}+z^t
\end{equation}
This equation sums the neighbouring node feature embeddings $f_t^\delta$ with the current node embedding $z^t$, where $\delta=\{\delta_1,\delta_2\}$, $t,\delta \in \{p,c,d\}$ such that $t \neq \delta$. Here $\odot$ is element-wise product and $\gamma_t$ are feature attention scores based on the current node embedding $z^t$ and the features embedding $f_t^\delta$: 
\begin{equation} \label{eq:gamma_t}
    \gamma_t=\sigma(W_t\cdot concat[z^t,f_t^\delta]+\sum\limits_{\delta \neq t}W_t^\delta f_t^\delta)
\end{equation}
where $W_t \in \mathbb{R}^{d \times 3d}$, $W_t^\delta \in \mathbb{R}^{d \times d}$ are learnable parameters.

In order to obtain feature embeddings $f_t^{\delta}$ for the given node $z^t$, we integrate homogeneous information from nodes with the same feature, as neighbouring nodes of the same feature can have varying levels of importance. The weights of each node for feature $\delta$ are defined as:
\begin{equation}\label{eq:mu_i}
    \mu_i=\frac{exp(\alpha_{\delta}^{\top} z_i^{\delta})}{\sum\limits_{z_i^{\delta} \in \mathcal{A}_t^{\delta}} exp(\alpha_\delta^{\top} z_i^{\delta})}
\end{equation}
where $\alpha_{\delta}^{\top}$ is the attention vector, and $\mathcal{A}_t^{\delta}$ represents each feature set. Then feature embeddings $f_t^\delta$ can be formulated as:
\begin{equation}\label{eq:p_t^delta}
    f_t^\delta=\sum_{i=1}^{\lvert \mathcal{A}_t^{\delta} \rvert} \mu_i z_i^{\delta}
\end{equation}

This module finally produces the semantics-enhanced feature embeddings, including price embedding set $H_p=[h_1^p,\ldots,h_{m_p}^p]$, the ID embedding set $H_d=[h_1^d,\ldots,h_{m_d}^d]$, and the category embedding set $H_c=[h_1^c,\ldots,h_{m_c}^c]$. For simplicity, we use $H_t=[h_1^t,\ldots, h_l^t]$ for subsequent operations.

\subsection{Basket-guided Dynamic Augmentation Networks}
The unified hybrid encoder captures complex global node relationships but struggles with dynamic user behavior. To address this, we introduce a basket-guided dynamic learning mechanism to enhance node embeddings. This approach maintains the global context while effectively learning local information, better adapting to individual basket sequences and user preferences.

Given that items within the same basket can reflect user preferences in general, we apply average pooling to aggregate their feature embeddings in each basket:
\begin{equation}\label{v_{B_i}^t}
    v_{B_i}^t=tanh(\frac{1}{l}\sum_{t=1}^l concat[h_1^t,..., h_l^t])
\end{equation}
Then we compile the feature vector 
$V^t=[v_{B_1}^t,  ..., v_{B_n}^t]$, where $n$ is the total number of baskets across all users in the batch.

After generating the basket-level representations, we identify the relevant baskets for a specific item. The presence of the same item in different baskets reflects its adaptability to various users needs or situations. Therefore, we construct item-to-basket correspondence matrices $\hat{V}_i^t=[v_{B_1}^t, ..., v_{B_j}^t]$, where $j$ represents the number of baskets in which item $i$ appears.

However, not all baskets containing the same item contribute equally to the recommendation task, some may contain irrelevant information or noise. To address this, we employ a self-attention mechanism to evaluate the impact of each basket and refine the overall preference for the item $i$.
\begin{equation}
    v_{\widetilde{i}}^t=\hat{V}_i^t \cdot \alpha^t
\end{equation}
\begin{equation}
    \alpha^t=softmax(b^{\top} \cdot tanh(W_{\alpha}\hat{V}_i^t))
\end{equation}
where $\alpha^t$ is attention weight vectors indicating the importance of each basket. $W_{\alpha}$ and $b$ are learnable parameters that adjust the model to the specific characteristics of the data. With the basket-guided dynamic augmentation network, we update the node embeddings as follows:
\begin{equation}
    \widetilde{h}_i^t=h_i^t+v_{\widetilde{i}}^t
\end{equation}

\subsection{User Behavior Modeling}
After obtaining the basket-augmented price embeddings $\widetilde{h}^p$ and ID embeddings $\widetilde{h}^d$, we build models to identify users' purchase decision-making behaviors (i.e., price sensitivities and product preferences). This approach enables the model to analyze diverse patterns in transaction data.

Using multi-head self-attention, we capture different attention patterns from the concatenated price embeddings $\widetilde{H}_p = concat[\widetilde{h}^p_1,\dots, \widetilde{h}^p_m]$ and weight matrices $W_n^Q, W_n^K, W_n^V \in \mathbb{R}^{d \times \frac{d}{h}}$ where $h$ is the number of heads, we generate the attention components:
\begin{equation}
    Q^n = \widetilde{H}_p W_n^Q, \quad 
    K^n = \widetilde{H}_p W_n^K, \quad 
    V^n = \widetilde{H}_p W_n^V
\end{equation}
We compute attention scores for each head and the final output $\varphi_p \in \mathbb{R}^{m \times d}$ is obtained by concatenating all heads:
\begin{equation}\label{eq:head_i}
    head_i = softmax (\frac{Q^n(K^n)^\top}{\sqrt{d/h}})V^n
\end{equation}
\begin{equation}\label{eq:varphi_p}
\varphi_p=concat[head_1,...,head_h]w^o
\end{equation}
where $w^o \in \mathbb{R}^{d \times d}$.

For product choice preferences, we utilize the Gated Linear Unit (GLU). We derive positional embeddings $E_{pos}$, reversed item embeddings $\widetilde{H}'_d$ \cite{desrosiers2010comprehensive}, and basket-level aggregated embeddings $H_{bask}$ that are derived from the original item embeddings $\widetilde{H}_d = concat[\widetilde{h}^d_1,\dots, \widetilde{h}^d_m]$:
\begin{equation}\label{eq:H_bask}
    H_{bask}=\frac{1}{L_b}\sum_{j=1}^{L_b}\widetilde{H}'_d[j]
\end{equation}
\begin{equation}\label{eq:GLU_N}          GLU\_N=GLU(E_{pos},\widetilde{H}'_d,H_{bask})
\end{equation}
where $L_b$ is the length of the basket. User's general interest embedding $\varphi_d$ is defined as:
\begin{equation}\label{eq:beta}
    \beta=W_{\beta} \cdot GLU\_N
\end{equation}
\begin{equation}\label{eq:varphi_d}  \varphi_d=\sum_{i=1}^{L_b}\beta^{(i)} \cdot \widetilde{H}'_d
\end{equation}
where $W_{\beta}$ is learnable parameter.
\subsection{Training Setup}
To compute the scores $y_i$ for all items $x_i \in I$, we perform an inner product between the user preference, learned from heterogeneous hypergraph and item features. We then apply a softmax function to derive the probabilities of each item being selected for the next basket:
\begin{equation}\label{eq:y_i}
    y_i=\varphi_p^\top z_i^p + \varphi_d^\top z_i^d
\end{equation}
\begin{equation}\label{eq:hat{y}}
    \hat{y}=softmax(y_i)
\end{equation}

The model is trained using a cross-entropy loss function, commonly employed in recommendation systems:
\begin{equation}\label{eq:mathcal{L_r}}
    \mathcal{L}_r=-\sum_{i=1}^n\alpha_i log(\hat{y}_i)+(1-\alpha_i)log(1-\hat{y}_i)
\end{equation}
where $\alpha$ is the one-hot encoding vector of the ground truth.

\section{Experiments}\label{Exp}
\subsection{Experiments Setup}

\textbf{Datasets.} We use two public datasets: Dunhumby\footnote{https://www.dunnhumby.com/source-files/} and Valuedshopper\footnote{https://www.kaggle.com/c/acquire-valued-shoppers-challenge/overview}, which are widely used for evaluating next basket recommendation algorithms. In Valuedshopper, we treated all items which were bought by same user in the same day as a basket, and then extracted 10k users sampled data. The characteristics of datasets after pre-processing is shown in the Table \ref{tab:stat_datasets}.

\begin{table}[hb!]
    \begin{center}
        \centering
        \caption{Statistical information of datasets.\label{tab:stat_datasets}}
        \begin{tabular}{lrr}
        \toprule    
        Dataset & Dunnhumby & Valuedshopper \\ 
        \midrule 
        \#Users & 2,462 & 9,992 \\
        \#Items & 12,150 & 6,739\\
        \#Price & 10 & 10 \\
        \#Categories & 235 & 490\\
        \#Baskets & 23,435 & 99,771\\
        Avg. items per basket & 9.6 & 9.7\\
        \bottomrule 
        \end{tabular}
    \end{center}
\end{table}

\textbf{Evaluation Metrics.} Following \cite{ariannezhad2023personalized,ariannezhad2022recanet, deng2023multi}, we choose the top K (i.e., K = 5, 10, 15) items in the ranking list of all items as the recommended set. To evaluate the performance of our model, we also adopt the widely used Normalized Discounted Cumulative Gain (NDCG) and Hit Rate (Hit).

\begin{table*}[h!]
        \centering
        \caption{Overall performance comparison between the baselines and our model. The bold is the best result and the underline is the second-best.\label{tab:results}}
        \begin{tabular}{lllllllll|l}
        \toprule    
        Dataset & Metric & FPMC & GRU4Rec & BERT4Rec & SASRec & CoHHN & TIFU-KNN & \textbf{BDHH} & Improve.\\ 
        \midrule 
        \multirow{6}*{Dunnhumby} & NDCG@5 & 0.0656 & 0.0525 & 0.0206 & 0.0620 & \underline{0.1084} & 0.0841 & \textbf{0.1570} & 44.8\%\\
        ~ & NDCG@10 & 0.0738 & 0.0597 & 0.0261 & 0.0699 & 0.0994 & \underline{0.1019} &  \textbf{0.1469} & 44.2\% \\
        ~ & NDCG@15 & 0.0788 & 0.0637 & 0.0279 & 0.0749 & 0.0980 & \underline{0.1127} &  \textbf{0.1428} & 26.7\% \\
        ~ & Hit@5 & 0.0881 & 0.0731 & 0.0300 & 0.0893 & 0.3199 & \underline{0.3387} &  \textbf{0.4109} & 21.3\%\\
        ~ & Hit@10 & 0.1133 & 0.0954 & 0.0471 & 0.1137 & 0.3967 & \underline{0.4320} &  \textbf{0.4758} & 10.1\% \\
        ~ & Hit@15 & 0.1324 & 0.1108 & 0.0540 & 0.1328 & 0.4442 & \underline{0.4787} &  \textbf{0.5059} & 5.7\% \\
        \midrule
        \multirow{6}*{Valuedshopper} & NDCG@5 & 0.0571 & 0.0703 & 0.0614 & 0.0675 & 0.0702 & \underline{0.1209} &  \textbf{0.1684} & 39.3\%\\
        ~ & NDCG@10 & 0.0702 & 0.0843 & 0.0745 & 0.0810 & 0.0847 & \underline{0.1516} &  \textbf{0.1637} & 8.0\% \\
        ~ & NDCG@15 & 0.0778 & 0.0928 & 0.0820 & 0.0890 & 0.0998 & \underline{0.1612} &  \textbf{0.1655} & 2.7\%\\
        ~ & Hit@5 & 0.0833 & 0.1002 & 0.0897 & 0.1008 & 0.1029 & \underline{0.4524} &  \textbf{0.4823} & 6.6\% \\
        ~ & Hit@10 & 0.1239 & 0.1439 & 0.1304 & 0.1427 & 0.1478 & \underline{0.5831} &  \textbf{0.5857} & 0.4\%\\
        ~ & Hit@15 & 0.1526 & 0.1760 & 0.1589 & 0.1729 & 0.2084 & \underline{0.6486} &  \textbf{0.6576} & 1.4\%\\
        \bottomrule 
        \end{tabular}
\end{table*}
\textbf{Baselines and Implementation Details.} We compared our model BDHH with the following recommendation models including both the conventional baselines and up-to-date deep methods that are often used for predicting items in recommender systems: FPMC \cite{rendle2010factorizing}, GRU4Rec \cite{hidasi2015session}, BERT4Rec \cite{sun2019bert4rec}, SASRec \cite{kang2018self}, CoHHN \cite{zhang2022price} and TIFU-KNN \cite{hu2020modeling}. Our code and dataset are available at https://github.com/HiYuening/BDHH.
The embedding size is set to 128, the learning rate is $1e-5$, the L2 regularizer is $1e-3$, the number of self-attention heads is 4, the batch size is set to 8 and 2 respectively, and all parameters in the model are optimized using the Adam optimizer.

\subsection{Performance Analysis}

We compare the proposed BDHH with baseline methods on two real-world datasets. The results are shown in Table \ref{tab:results}. We can observe that: (1) Models that focus on modeling sequential patterns, including SASRec, BERT4Rec, and GRU4Rec, are the weakest baselines on the Dunnhumby dataset because they only consider the sequential relationships of adjacent items and ignore the complex relationships between items within the basket. Specifically, SASRec outperforms GRU4Rec because they use self-attention mechanisms to capture long-term dependencies in user historical behaviors. In contrast, GRU4Rec uses gated recurrent units (GRUs) to focus on short-term dependencies of sequences. In particular, comparing the experimental results on Dunnhumby and Valuedshopper, we find that SASRec and GRU4Rec perform better than BERT4Rec. (2) CoHHN outperforms SASRec, BERT4Rec, GRU4Rec and FPMC because CoHHN constructs a hypergraph network to learn the impacts of item price. Moreover, FPMC is relatively competitive because it considers adjacent basket relationships. TIFU-KNN performs the best among baseline models on both datasets, demonstrate that the importance of incorporating within-basket items correlation. (3) Our proposed BDHH achieves the best performance in all datasets, with relative improvements ranging from 0.4\% to 44.8\%. Based on the comparison with the above baselines, we attribute the improvement to two key factors: first, we introduce multiple features of items to build a multi-feature heterogeneous hypergraph network to accurately model the universal representation of items, and second, we build a basket-guided dynamic enhancement network by considering the correlation between baskets and items to enhance the representation of item features in order to more effectively model user preferences.

\subsection{Ablation Study}
In this section, we analyze how different components of BDHH affect the final performance. We compare two ablation variants, including (1) $\underline{w/o\hspace{0.1cm}\mathrm{A}}$ removing the basket-guided dynamic augmentation module, and (2) $\underline{w/o\hspace{0.1cm}\mathrm{P}}$ removing price-based user behavior model. The experimental results of the proposed BDHH and its variants are shown in Table \ref{tab: Ablation}. We can observe that each component is effective in significantly improving performance. Notably, $\underline{w/o\hspace{0.1cm}\mathrm{A}}$ has the worst performance across both datasets and metrics except NDCG, which highlights the importance of learning item-basket-user relationships. Besides, $\underline{w/o\hspace{0.1cm}\mathrm{P}}$ shows a significant performance drop, ranging from 12.7\% to 28.3\%, confirming the crucial role of price information in capturing user behavior patterns and validating the effectiveness of price-based user preference modeling.


\begin{table}[h!]
    \centering
    \caption{Ablation study of BDHH variants on Dunnhumby and ValueShopper datasets with K=10.}\label{tab: Ablation}
    \begin{tabular}{lcccc}
        \toprule
        \multirow{2}{*}{Methods} & \multicolumn{2}{c}{Dunhumby} & \multicolumn{2}{c}{Valuedshopper} \\
        \cmidrule(r){2-3} \cmidrule(r){4-5}
        & NDCG  & Hit & NDCG  & Hit \\
        \midrule
        BDHH    & 0.1469  & 0.4758 & 0.1637   &  0.5857 \\
        $\underline{w/o\hspace{0.1cm}\mathrm{A}}$  & 0.1123  & 0.4044 & 0.1210  & 0.5071 \\
        $\underline{w/o\hspace{0.1cm}\mathrm{P}}$ & 0.1053  & 0.4084 & 0.1235  & 0.5111\\
        \bottomrule
    \end{tabular}
\end{table}


\section{Conclusions and Future Work}\label{Conclu}
In this work, we propose a novel NBR model Basket-augmented Dynamic Heterogeneous Hypergraph (BDHH), which employs a heterogeneous hypergraph to explore the intricate relationships within the recommendations process, and develop a dynamic basket-augmentation network to strengthen the connections among items, baskets and users, incorporating price and other essential item factors. Experimental results on real-world datasets demonstrated that our model outperforms competitive baselines and could provide more accurate item combinations prediction. Furthermore, we have studied the contribution of each component in our model through an ablation study. This study confirmed that all components, specifically designed to address the identified research gaps, are essential for achieving optimal performance. In the future, we aim to investigate deeper correlations among item features and to explore the broader influence of price on users’ purchase decisions.

\newpage
\bibliographystyle{IEEEtran}
\bibliography{reference}{}

\vspace{12pt}

\end{document}